\documentclass[11pt,document,nofootinbib,superscriptaddress,onecolumn,preprintnumbers,balancelastpage]{article}

\usepackage{cancel}

\usepackage{framed}


\usepackage{amsmath, amsfonts, amsthm, amssymb}
\usepackage[normalem]{ulem}
\usepackage{cite}

\usepackage[toc,page]{appendix}

\usepackage{bm}
\usepackage{upgreek}
\usepackage{comment}
\usepackage{ifpdf}
\usepackage{xstring}
\usepackage{setspace}
\usepackage[utf8]{inputenc}
\usepackage{hyperref}
\usepackage{graphicx}  
\usepackage{breqn}
\usepackage{mathrsfs}

\DeclareMathAlphabet{\mathbbold}{U}{bbold}{m}{n}

\newcount\Comments  
\usepackage{color}
\definecolor{colorGH}{rgb}{.2,.7,.2}
\newcommand{\kibitz}[2]{\ifnum\Comments=1\textcolor{#1}{#2}\fi}

\let\isout\sout \renewcommand{\sout}[1]{\ifmmode\text{\isout{\ensuremath{#1}}}\else\isout{#1}\fi}

\newcommand{\be}{\begin{equation}}
\newcommand{\ee}{\end{equation}}
\newcommand{\bea}{\begin{eqnarray}}
\newcommand{\eea}{\end{eqnarray}}

\newcommand{\eq}[2][]{%
    \ifx&#1&
      \begin{equation*} #2 \end{equation*}
    \else
  \begin{equation}\label{#1}{#2} \end{equation}
\fi
}
\newcommand{\eqs}[2][]{%
    \ifx&#1&
      \begin{eqnarray*} #2 \end{eqnarray*}
    \else
  \begin{subequations}\label{#1}\begin{eqnarray} #2 \end{eqnarray}\end{subequations}
\fi
}

\makeatletter
\renewcommand*{\p@section}{\S\,}
\renewcommand*{\p@subsection}{\S\,}
\makeatother
\begin{document}

\begin{center}\Large{Double Kerr-Schild spacetimes and the Newman-Penrose map}
\end{center}




\thispagestyle{empty}

\begin{center}
Kara Farnsworth\textsuperscript{1},
Michael L. Graesser\textsuperscript{2},
Gabriel Herczeg\textsuperscript{3}
\end{center}

\begin{center}
{\bf 1} {\footnotesize D\'{e}partment de Physique Th\'{e}orique,
Universit\'{e} de Gen\`{e}ve, CH-1211 Gen\`{e}ve, Switzerland}
\\
{\bf 2} {\footnotesize Theoretical Division, Los Alamos National Laboratory, Los Alamos, NM 87545, USA}
\\
{\bf 3} {\footnotesize Brown Theoretical Physics Center, Department of Physics, Brown University, Providence, RI 02912, USA}
\\
\end{center}

\begin{center}
\today
\end{center}

\begin{abstract}
\noindent
The Newman-Penrose map, which is closely related to the classical double copy, associates certain exact solutions of Einstein's equations with self-dual solutions of the vacuum Maxwell equations. Here we initiate an extension of the Newman-Penrose map to a broader class of spacetimes. As an example, we apply principles from the Newman-Penrose map to associate a self-dual gauge field to the Kerr-Taub-NUT-(A)dS spacetime and we show that the result agrees with previously studied examples of  classical double copies.
The corresponding field strength
exhibits a discrete electric-magnetic duality that is distinct from its (Hodge star) self-dual property.
\end{abstract}

\tableofcontents

\newpage
\section{Introduction}
\label{sec:intro}
The double copy is a family of correspondences between gravitational and gauge theories with origins in string theory \cite{Kawai:1985xq} that can be summarized schematically as
\begin{align*}
\textrm{gravity} = \textrm{gauge}\otimes \textrm{gauge}.
\end{align*}
Years after the initial discovery that closed string amplitudes can be factorized into a product of open string amplitudes, a new manifestation of the double copy was discovered in the form of color-kinematics duality, where a kinematic identity analogous to the Jacobi identity for color factors was introduced to argue that the Kawai-Lewellen-Tye (KLT) relations of \cite{Kawai:1985xq} are equivalent to a ``diagram-by-diagram numerator `squaring' relation with gauge theory"  \cite{Bern:2008qj}. This relationship greatly simplifies calculations in gravity, and has provided useful insight into gauge theory calculations. It has been proven at tree level for Yang-Mills theory and Einstein gravity amplitudes \cite{Bern:2010yg}, and in various limits and cases with extra symmetry at higher loop level \cite{Bern:2010ue, BjerrumBohr:2012mg, Nohle:2013bfa, Boels:2013bi, Bern:2013yya, Du:2014uua}, however a general exact equivalence at all orders has yet to be found. For a recent review on the subject of the double copy, see \cite{Adamo:2022dcm}.

In an effort to understand how general this double copy relationship is, a proliferation of similar approaches has been initiated in recent years, particularly relating \emph{exact} non-perturbative classical solutions of gauge theory and gravity
\cite{Monteiro:2014cda, Elor:2020nqe, Farnsworth:2021wvs, Luna:2015paa, Luna:2016due, White:2016jzc,  Goldberger:2016iau, Adamo:2017nia, DeSmet:2017rve, Bahjat-Abbas:2017htu, Carrillo-Gonzalez:2017iyj, Lee:2018gxc,Berman:2018hwd,Gurses:2018ckx, Luna:2018dpt, Bahjat-Abbas:2018vgo, CarrilloGonzalez:2019gof, Cho:2019ype, Arkani-Hamed:2019ymq, Bah:2019sda, Huang:2019cja, Alawadhi:2019urr, Kim:2019jwm, Banerjee:2019saj, Bahjat-Abbas:2020cyb, Luna:2020adi, Keeler:2020rcv,  Alawadhi:2020jrv, Easson:2020esh, Chacon:2020fmr, delaCruz:2020bbn, Godazgar:2020zbv, Emond:2020lwi, Berman:2020xvs, Cheung:2020djz, Prabhu:2020avf, White:2020sfn, Monteiro:2020plf, Guevara:2020xjx, Campiglia:2021srh,Alkac:2021bav, Chacon:2021wbr, Chen:2021chy, Alkac:2021seh,  Angus:2021zhy, Gonzalez:2021bes, Chacon:2021hfe, Gonzo:2021drq,Cho:2021nim,Adamo:2021dfg,Godazgar:2021iae,Mao:2021kxq,Moynihan:2021rwh,  Easson:2021asd,Guevara:2021yud,Chacon:2021lox,Monteiro:2021ztt,Gonzalez:2021ztm, Ben-Shahar:2021zww, CarrilloGonzalez:2022mxx,Emond:2022uaf, Kosower:2022yvp, Han:2022ubu, Armstrong-Williams:2022apo,Alkac:2022tvc, Han:2022mze,  Luna:2022dxo, Chawla:2022ogv,Didenko:2022qxq,Easson:2022zoh,Nagy:2022xxs,Dempsey:2022sls,CarrilloGonzalez:2022ggn, Ben-Shahar:2022ixa,Armstrong-Williams:2023ssz,Chawla:2023bsu, Bonezzi:2023pox, Szabo:2023cmv, Easson:2023dbk, Brown:2023zxm, Borsten:2023paw, Ball:2023xnr}. The first classical ``Kerr-Schild" double copy was introduced in \cite{Monteiro:2014cda}, where stationary, vacuum solutions of Einstein's equations of Kerr-Schild type were mapped to solutions of the vacuum Maxwell's equations. 
Since then, many other classical double copies have been proposed, including one called the Newman-Penrose map, which we introduced in \cite{Elor:2020nqe} and elaborated upon in \cite{Farnsworth:2021wvs} using the language of twistors. 

The Newman-Penrose map associates a self-dual solution of the vacuum Maxwell equations with a solution of the Einstein equations in Kerr-Schild form. Unlike other versions of the double copy, including the Kerr-Schild double copy, this map generically applies to gravitational solutions that are not necessarily stationary or vacuum. When  other versions of the double copy and the Newman-Penrose map are both defined, they agree in all known examples, although a general proof of their equivalence has not been conclusively established. 

In this note, we apply concepts from the Newman-Penrose map in an exploratory manner to a family of spacetimes for which the Newman-Penrose map is not well defined a priori. In particular, we consider the Kerr-Taub-NUT-(A)dS family, which is Petrov type D, and which can be given a double Kerr-Schild form after analytic continuation to Kleinian signature \cite{Chong:2004hw}. A straightforward application of the Newman-Penrose map to each of the null, geodesic and shear-free Kerr-Schild vectors leads to a complex self-dual gauge field whose real part agrees with previous double copy studies of the same metric. The resulting field strength is shown to exhibit 
an electric-magnetic duality, that acts to relate the real (imaginary) parts of the electric and magnetic fields to each other, and consequently is a symmetry of the solution that is distinct from its (Hodge star) self-dual property.


\section{Newman-Penrose map for single Kerr-Schild spacetimes}
\label{sec:NP-map-Single-KS}

In this section we give a concise overview of the Newman-Penrose map \cite{ Farnsworth:2021wvs, Elor:2020nqe} for single Kerr-Schild spacetimes in a general metric signature. We focus on the basic ingredients of the map that must be generalized in order to accommodate more complicated spacetimes.  

The Newman-Penrose map consists of two main components:
\begin{itemize}
\item $\Phi$: a scalar function, harmonic with respect to the background metric, which arises in a particular choice of null tetrad that is suitable for a generic\footnote{By `generic,' we mean that a tetrad of the form \eqref{eq:tetrad} exists whenever the Kerr-Schild vector is shear-free, geodesic and expanding. This includes nearly all known vacuum Kerr-Schild spacetimes with the exception of the PP wave spacetimes which are Petrov type N and have a non-expanding congruence.} Kerr-Schild spacetime.
\item $\hat{k}$: a differential spin-raising operator. 
\end{itemize}
From these elements we can construct the gauge field
\begin{align}
\label{eq:Adef}
A = \hat{k}\Phi.
\end{align}
When $\Phi$ is harmonic with respect to the flat background metric, the field strength $F=dA$ defined by the gauge field defined in \eqref{eq:Adef} is self-dual, and as a consequence automatically satisfies the vacuum Maxwell equations in flat spacetime.

The original formulation of the Newman-Penrose map applies to single Kerr-Schild spacetimes with a flat background
\begin{align}
g_{\mu\nu} = g^{(0)}_{\mu\nu} + V \ell_\mu \ell_\nu, 
\end{align}
where $\ell_\mu$ is taken to be null with respect to both the full ($g_{\mu\nu}$) and background ($g^{(0)}_{\mu\nu}$) metrics. In Lorentzian signature, if the stress tensor satisfies $T_{\mu\nu}\ell^\mu \ell^\nu = 0$ (which includes vacuum solutions of Einstein's equations with or without a cosmological constant), then $\ell^\mu$ is also geodesic with respect to both the full and background metrics (see theorem (32.1) of \cite{Stephani:2003tm}). Moreover, when the vacuum Einstein equations are satisfied, it can be shown that $g_{\mu \nu}$ is algebraically special and $\ell^\mu$ is a repeated principle null direction of the Weyl tensor \cite{Chandrasekhar:1985kt},
 so the Goldberg Sachs theorem ensures $\ell^\mu$ is also shear-free \cite{2009GReGr..41..433G}.  Here, we wish to consider more general spacetimes, which may have an arbitrary metric signature, or even be complex, and which may be sourced by more general stress tensors. However, we will nevertheless assume that $\ell^\mu$ is geodesic and shear-free. 
Any such spacetime admits a null tetrad of the form 
\begin{align}
\begin{split}
\ell &=  du+ \tilde{\Phi}d\zeta + \Phi d\tilde{\zeta} + \Phi\tilde{\Phi}dv\, , \\
n &= dv + \tfrac{1}{2}V \ell ,\\
m &=   -(\tilde{\Phi} dv + d\tilde{\zeta})\\
\tilde{m} &=   -(\Phi dv + d\zeta) 
\label{eq:tetrad}
\end{split}
\end{align}
where $\ell = \ell_\mu dx^\mu$, etc. $u, v, \zeta, \tilde{\zeta}$ are null coordinates\footnote{In Kleinian signature (which will be the focus of the paper) these null coordinates are defined in terms of Cartesian and twisted spheroidal coordinates by
\begin{align}
\begin{split}
\label{eq:flatcoor}
u &= \tfrac{1}{\sqrt{2}}(t - z) = \frac{1}{\sqrt{2}}(t - r\cos\theta) \\ 
v &= \tfrac{1}{\sqrt{2}}(t+ z) = \frac{1}{\sqrt{2}}(t + r\cos\theta) \\ \quad \zeta &= \tfrac{1}{\sqrt{2}}(x + y) = \frac{1}{\sqrt{2}}(r+a)e^{\varphi} \sin\theta\\
\tilde{\zeta} &= \tfrac{1}{\sqrt{2}}(x - y) = \frac{1}{\sqrt{2}}(r-a)e^{-\varphi} \sin\theta.
\end{split}
\end{align}}, and $V$, $\Phi$ and $\tilde{\Phi}$ are scalar functions. In Lorentzian signature, the tilded and untilded quantities are related by complex conjugation, however in general signature they are independent.

The full metric can then be written
\begin{align} 
g_{\mu\nu} &= n_\mu \ell_\nu + \ell_\mu n_\nu - m_\mu \tilde{m}_\nu - \tilde{m}_\mu m_\nu \\
&= g^{(0)}_{\mu\nu} + V\ell_\mu \ell_\nu
\end{align} 
with $g^{(0)}_{\mu\nu}dx^\mu dx^\nu = 2(dudv - d\zeta d\tilde{\zeta})$.  
For vacuum spacetimes, $V$ is determined by $\Phi$ and $\tilde{\Phi}$ and the mass $M$, so $\Phi$ and $\tilde{\Phi}$ completely determine a vacuum solution up to the constant $M$.

The shear-free and geodesic conditions imply that $\Phi$ satisfies the following non-linear partial differential equations:
\be
\left(\Phi \partial_\zeta  - \partial_v\right) \Phi = 0, \qquad
\big(\Phi \partial_u  - \partial_{\tilde{\zeta}}\big) \Phi = 0. \label{Non-linear}
\ee

Importantly, these equations also imply that $\Phi$ is harmonic with respect to the flat background metric
\begin{align}
\Box_0 \Phi = (\partial_u \partial_v - \partial_\zeta \partial_{\tilde{\zeta}}) \Phi = 0.
\end{align}
This is also the (linearized) equation of motion for the zeroth copy biadjoint scalar field. Pushing this intuition further, we can construct a gauge field by acting on $\Phi$ with the spin-raising operator
\be
\hat{k} = -\frac{Q}{2\pi\epsilon_0}(dv\,\partial_\zeta + d\tilde{\zeta}\,\partial_u).\label{diffOp}
\ee
 The gauge field defined by
\begin{align}
A_{\textnormal{NP}} = \hat{k}\Phi
\end{align}
automatically satisfies the self-dual vacuum Maxwell equations when $\Phi$ is harmonic with respect to the flat background metric. It is worth noting that the operator $\hat{k}$ used above in defining the Newman-Penrose map also appears in the definition of the self-dual Kerr-Schild double copy originally presented in \cite{Monteiro:2014cda} and further developed in subsequent works. However, here $\Phi$ satisfies the non-linear PDEs \eqref{Non-linear}, but we do not impose the Pleba\'{n}ski equation. Consequently, $h_{\mu\nu} = \hat{k}_\mu\hat{k}_\nu\Phi$ need not correspond to a self-dual solution of Einstein's equations. With this construction, we can define a map from any single Kerr-Schild metric with an expanding, shear-free, null geodesic congruence to a self-dual solution of the flat space vacuum Maxwell equations.  This includes all vacuum Kerr-Schild spacetimes with an expanding repeated principle direction. However, unlike the Kerr-Schild double copy, the map also applies to non-vacuum and non-stationary spacetimes obeying these conditions. While there is no general proof of the equivalence of the Newman-Penrose map and the Kerr-Schild double copy, they have been shown to be equivalent where both formalisms apply \cite{Elor:2020nqe}.

\section{Newman-Penrose map for double Kerr-Schild spacetimes}
To explore the generality of this framework, we attempt to apply the map to both double Kerr-Schild spacetimes and solutions in non-flat (maximally symmetric) backgrounds:
\begin{align}
g_{\mu\nu} = \bar{g}_{\mu\nu} + \phi k_\mu k_\nu + \psi \ell_\mu \ell_\nu.\label{DKS}
\end{align}
 
Here $\bar{g}_{\mu\nu}$ is maximally symmetric, and we will write it in a manifestly conformally flat form for ease of comparison with the single Kerr-Schild case. This metric is in double Kerr-Schild form, so we have two scalar fields $\phi$ and $\psi$, as well as two vectors $k_\mu$ and $\ell_\mu$ which are assumed to each be null, geodesic, and shear-free with respect to the background metric, from which it follows that they are null, shear-free and geodesic with respect to the full metric. The vectors $k_\mu$ and $\ell_\mu$ are also assumed to be orthogonal, which cannot be accomplished for real metrics in Lorentzian signature. Hence, our discussion of double Kerr-Schild metrics will be primarily in Kleinian signature unless otherwise noted, although it applies equally well to complex spacetimes where metric signature is not a well-defined notion. 

In Kleinian signature, the background metric (which can be de Sitter (dS), anti-de Sitter (AdS), or flat) can be written in null coordinates as \cite{Griffiths:2009dfa}
\begin{align}
\bar{g}_{\mu\nu} dx^\mu dx^\nu =\frac{2}{\Delta^2}\left( dU dV - dX dY\right)
\end{align}
with $\Delta = 1 - \frac{\lambda}{2}(UV - XY)$ and the sign of $\lambda$ corresponding to dS (positive), AdS (negative) or zero (flat). In this signature, the coordinates $X$ and $Y$ are real and are not complex conjugates of one another, but they become conjugate upon analytic continuation back to Lorentzian signature. 

 In these coordinates, our null vectors can be normalized so that they take the form
\begin{align}
k &= dU + \tilde{\Phi}_k\, dX + \Phi_k\, dY + \Phi_k\, \tilde{\Phi}_k\, dV \\
\ell &= dU + \tilde{\Phi}_\ell\, dX + \Phi_\ell\, dY + \Phi_\ell\, \tilde{\Phi}_\ell\, dV.
\end{align}
 In analogy with the single Kerr-Schild case, the (real) scalar functions $\Phi_k$ and $\Phi_\ell$ will be used to construct the gauge fields associated with the vectors $k$ and $\ell$.
 
 We now also have two spin-raising operators
\begin{align}
\label{eq:khatem}
\hat{k}_e = \frac{-Q_e}{2\pi \epsilon_0} \left( dV \partial_X + dY \partial_U\right), \quad{}
\hat{k}_m = \frac{-Q_m}{2\pi \mu_0} \left( dV \partial_X + dY \partial_U\right).
\end{align}
The operator $\hat{k}_e$ is  identical to $\hat{k}$ for single Kerr-Schild metrics and the operator $\hat{k}_m$ is the magnetic analog. The total gauge field obtained from this double Kerr-Schild metric via the Newman-Penrose map is then the linear combination
\begin{align}
\label{eq:Adoubledef}
A_{\textnormal{NP}} &= \hat{k}_e \Phi_\ell + \hat{k}_m \Phi_k.
\end{align}
 As was the case for single Kerr-Schild metrics, the self-dual gauge field \eqref{eq:Adoubledef} satisfies the vacuum Maxwell equations by construction, provided both $\Phi_k$ and $\Phi_\ell$ are harmonic with respect to the flat metric. Note that since the wave operator is not conformally invariant, $\Phi_k$ and $\Phi_\ell$ are \emph{not} solutions of the wave equation on an (A)dS background. However, since the vacuum Maxwell equations \emph{are} conformally invariant, the gauge field \eqref{eq:Adoubledef} is a solution on both the flat and maximally symmetric backgrounds. Generalizing this formalism to double Kerr-Schild metrics and maximally symmetric backgrounds then amounts to understanding when this harmonic condition holds. 

 Although we are not able to prove that the functions $\Phi_\ell$ and $\Phi_k$ are harmonic in general double Kerr-Schild spacetimes, for the Kerr-Taub-NUT-(A)dS family of solutions, the Newman-Penrose map outlined here can be constructed explicitly in the affirmative. This is the subject of the proceeding section, and here we state the main results. The function $\Phi_\ell$ is shown to satisfy the following non-linear PDEs 
\begin{align} 
\begin{split}
 \label{eqn:Klein-space-non-linear-PDEs}
\left(\Phi_\ell \partial_X 
 - \partial_V \right) \Phi_\ell =0,\quad{} \left(\Phi_\ell \partial_U -\partial_Y \right)\Phi_\ell = 0
\end{split}
\end{align} 
which are the same as the non-linear PDEs \eqref{Non-linear} but for the null coordinates of the curved spacetime. This result is not surprising, but also non-trivial, at least for these coordinates. It is not surprising because the (A)dS background is conformally flat, and the geodesic and shear-free properties of $k$ and $\ell$ are conformally invariant. Hence, both essential properties of $k$ and $\ell$ required to define the Newman-Penrose map are preserved in passing to a maximally symmetric background.  On the other hand, the explicit form for $\Phi_\ell$ has a complicated dependence on both the cosmological constant and null coordinates, such that satisfying (\ref{eqn:Klein-space-non-linear-PDEs}) is not trivial.

It follows from 
(\ref{eqn:Klein-space-non-linear-PDEs})
that $\Phi_\ell$ is harmonic with respect to the flat metric, and consequently the field strength obtained from the gauge potential $A_\ell =\hat{k}\Phi_\ell$ is automatically self-dual and satisfies the vacuum Maxwell equations in both flat and conformally flat spacetimes. The real part of the gauge field $\Re(A_\ell)$ is shown to be gauge-equivalent to the original double copy prescription for obtaining a gauge field from the Kerr-Schild form of the metric.

As shown below, a similar story holds for $\tilde{\Phi}_\ell$, $\tilde{\Phi}_k$, and $\Phi_k$ - they satisfy similar non-linear PDEs and are therefore each harmonic with respect to the flat background metric. For double Kerr-Schild spacetimes, {\em a priori} there could exist multiple Newman-Penrose maps using different combinations of $\Phi_\ell$, $\tilde{\Phi}_\ell$, $\tilde{\Phi}_k$, and $\Phi_k$, however for this family of spacetimes these four functions are sufficiently related to one other that the possible different self-dual gauge fields obtained from acting with $\hat{k}_e$ and $\hat{k}_m$ are  equivalent.

\section{Kerr-Taub-NUT-(A)dS metric}
\label{sec:Kerr-Taub-NUT-(A)dS-metric}

A particular solution that admits a double Kerr-Schild form containing many well known solutions as various limits is the Kerr-Taub-NUT-(A)dS metric. This solution can be written in a double Kerr-Schild form that is real-valued in $(2,2)$ signature after analytic continuation. 

In Plebanski coordinates it is given as \cite{Chong:2004hw} 
\begin{align}
ds^2 = d\bar{s}^2 -\frac{2Mq}{q^2-p^2} \left[d\tilde{\tau} + p^2 d\tilde{\sigma}\right]^2- \frac{2N p}{q^2-p^2} \left[d\tilde{\tau} + q^2 d \tilde{\sigma} \right]^2.\label{KerrTaubNUTAdS}
\end{align}
Here $M$ is the Schwarzschild mass, $N$ is the NUT charge, and
\begin{align}
d\bar{s}^2 = & \frac{\bar{\Delta}_p}{q^2-p^2}\left[d\tilde{\tau} + q^2 d\tilde{\sigma}\right]^2 +\frac{\bar{\Delta}_q}{q^2-p^2} \left[ d\tilde{\tau} + p^2 d\tilde{\sigma}\right]^2 +2\left[d\tilde{\tau} + q^2 d\tilde{\sigma}\right] dp+2\left[d\tilde{\tau} + p^2 d\tilde{\sigma}\right] dq
\end{align}
is the background (A)dS metric.
 Here $\bar{\Delta}_p = \gamma - \epsilon p^2 + \lambda p^4$  and $\bar{\Delta}_q= -\gamma + \epsilon q^2 - \lambda q^4$ 
 with $\lambda $ related to the scalar curvature via $R =  12 \lambda$, and $\gamma$ and $\epsilon$ will be related to the rotational parameter $a$. 

 In these coordinates the two orthogonal Kerr-Schild vectors take especially simple forms,
 \begin{align} 
 \ell &= d \tilde{\tau} + p^2 d \tilde{\sigma} \\
 k &= d \tilde{\tau} + q^2 d \tilde{\sigma} 
 \end{align}
 which hereafter are referred to as the ``mass'' and ``NUT'' vectors, respectively.
 
To guide our intuition, we first transform this to twisted spheroidal coordinates using
\begin{align}
\begin{split}
dq &= dr\\
dp &= d(a \cos\theta)\\
d\tilde{\tau} &= \frac{1}{(1-\lambda a^2)}\left(dt-\frac{1}{(1-\lambda r^2)}dr +ad\varphi\right) - \frac{a \cos^2\theta}{(1-\lambda a^2 \cos^2\theta)\sin\theta} d\theta \\
d\tilde{\sigma} &= \frac{1}{(1-\lambda a^2)}\left(-\lambda dt+ \frac{\lambda}{(1-\lambda r^2)}dr - \frac{1}{a}d\varphi\right)+ \frac{1}{(1-\lambda a^2 \cos^2\theta)a\sin\theta} d\theta.
\end{split}
\label{twistedspheroidalpq}
\end{align}
With the choices  $\gamma = a^2$ and $\epsilon  = 1 + \lambda a^2$ the  background metric becomes
\begin{align}
\begin{split}
d\bar{s}^2 =&\ \frac{(1-\lambda r^2)(1-\lambda a^2 \cos^2\theta) }{(1-\lambda a^2) }dt^2-\frac{(1-\lambda a^2\cos^2\theta)}{ (1-\lambda r^2)(1-\lambda a^2)} dr^2 \\
&- \frac{(r^2-a^2\cos^2\theta)}{(1-\lambda a^2\cos^2\theta)} d\theta^2+ \frac{(r^2-a^2)\sin^2\theta }{(1-\lambda a^2)}d\varphi ^2 +2 \frac{a\sin^2\theta }{(1-\lambda a^2)}d\varphi  dr.
\end{split}
\label{metric:background}
\end{align} 
After analytic continuation to Lorentzian signature ($a\rightarrow - ia, \ \varphi \rightarrow i \varphi$), this metric reduces to known results in particular limits:  taking $\lambda \rightarrow 0$ gives the Minkowski metric in twisted spheroidal coordinates used in \cite{Elor:2020nqe}, and taking $a\rightarrow 0$ gives the (A)dS metric in the familiar static spheroidal coordinates. The parameter $a$ is associated with the angular momentum of the black hole, as in the case of the Kerr metric. In these coordinates, the  mass vector $\ell$ and NUT vector $k$ become
\begin{align}
\ell &=  \frac{(1-\lambda a^2\cos^2\theta)}{(1-\lambda a^2)}dt -\frac{(1-\lambda a^2\cos^2\theta)}{(1-\lambda r^2)(1-\lambda a^2)} dr +\frac{a\sin^2\theta}{(1-\lambda a^2)}d\varphi\label{nullvectorell}\\
k &= \frac{(1- \lambda r^2)}{(1- \lambda a^2)} dt - 
 \frac{1}{(1- \lambda a^2)} dr + \frac{(r^2-a^2 \cos^2 \theta)}{a \sin \theta 
 (1- \lambda a^2 \cos^2 \theta)} d \theta - \frac{(r^2-a^2)}{a(1-\lambda a^2)}d \varphi.
 \label{nullvectork}
\end{align}
Once again we see that in the limit $\lambda \rightarrow 0$, $\ell$ matches with the Schwarzschild ($a = 0$) or Kerr ($a \neq 0$) vectors used in \cite{Elor:2020nqe} after analytic continuation. For the vector $k$, the $a\rightarrow 0$ limit is singular, however the functions $\Phi_k$ and $\tilde{\Phi}_k$ are not, as shown below.

\subsection{Scalar functions}
In order to extract the scalar functions analogous to $\Phi$ in the single Kerr-Schild case, we perform a second coordinate transformation 
\begin{align}
\begin{split}
U &= \frac{\Delta}{\sqrt{2} }\bigg[ \frac{1}{\sqrt{\lambda}} \frac{\sqrt{1-\lambda r^2}}{\sqrt{1-\lambda a^2}}\sqrt{1-\lambda a^2\cos^2\theta} \sinh(\sqrt{\lambda} t) - r\cos\theta\bigg] \\
V &= \frac{\Delta}{\sqrt{2}}\bigg[ \frac{1}{\sqrt{\lambda}} \frac{\sqrt{1-\lambda r^2}}{\sqrt{1-\lambda a^2}}\sqrt{1-\lambda a^2\cos^2\theta} \sinh(\sqrt{\lambda} t) + r\cos\theta\bigg]  \\
X &= \frac{\Delta}{\sqrt{2}}\frac{1}{\sqrt{(1-\lambda a^2)}}(r-a) e^{\varphi}\sin\theta \\
Y &= \frac{\Delta}{\sqrt{2}}\frac{1}{\sqrt{(1-\lambda a^2)}}(r+a) e^{-\varphi}\sin\theta 
\end{split}
\end{align}
with $\Delta = 1 - \frac{\lambda}{2}(UV - XY)$ and $r$ defined by the equation
\begin{align}
     \frac{(V-U)^2}{2r^2} + \frac{2XY}{r^2-a^2}(1-\lambda a^2) = \Delta^2.
 \label{eqn:maximally-symmetric-space-r-relation}
 \end{align}

The limit $\lambda \rightarrow 0$ recovers 
the flat spacetime transformation to null coordinates \eqref{eq:flatcoor}. In these null coordinates, the maximally symmetric background metric becomes
\begin{align}
d\bar{s}^2 = \frac{2}{\Delta^2} \left( dU dV - dX dY\right)
\end{align}
which is conformally flat and reduces to the flat space metric in null coordinates for $\lambda \rightarrow 0$.

The mass and NUT vectors written in these coordinates are
\begin{align}
\ell
=&\ \mathcal{N}_\ell\left[ dU+\tilde{\Phi}_\ell dX + \Phi_\ell dY + \Phi_\ell \tilde{\Phi}_\ell dV\right] \\
k
=&\   \mathcal{N}_k \left[dU+\tilde{\Phi}_k dX + \Phi_k dY + \Phi_k \tilde{\Phi}_k d V\right]
\label{eqn:(A)dS-Taub-NUT-null-vectors-l-and-k}
\end{align}
with normalization factors
\begin{align}
\mathcal{N}_\ell  &= \frac{ (1+\cos\theta)}{2\Delta(1-\lambda r^2)}\left[\sqrt{2}-\lambda r V  - \frac{\lambda\Delta}{\sqrt{2}} (1-\cos\theta)\frac{ (r^2- a^2)}{1-\lambda a^2} \right]\\
\mathcal{N}_k &= \frac{  (r+a)  }{2a\Delta (1-\lambda a^2\cos^2\theta)}\left(\sqrt{2}  - \lambda a  V \cos\theta-\frac{\lambda\Delta}{\sqrt{2}}  (1-\cos^2\theta)  \frac{a(r-a)}{ (1-\lambda a^2)} \right)
\end{align}
and scalar functions 
\begin{align}
\Phi_\ell &= \frac{\left[\frac{1}{\sqrt{2} } \lambda(r-a) (U+V+\sqrt{2}\Delta r)- 2(1-\lambda a r)\right]}{2\mathcal{N}_\ell \Delta^2 (r-a)(1-\lambda r^2)} X\label{eqn:general-Phi-ell}\\
\tilde{\Phi}_\ell &= \frac{\left[\frac{1}{\sqrt{2} } \lambda (r+a)(U+V+\sqrt{2}\Delta r)-2(1+\lambda a r)\right]}{2\mathcal{N}_\ell  \Delta^2 (r+a)(1-\lambda r^2)}  Y\label{eqn:general-tilde-Phi-ell}
\end{align}

\begin{align}
\Phi_k&= \frac{\left[\frac{1}{\sqrt{2}} \lambda a(1-\cos\theta) \left(U+V+ \sqrt{2}\Delta  a\cos\theta\right)+ 2  (1 -\lambda a^2\cos\theta) \right]}{2\mathcal{N}_k  \Delta^2 a (1-\lambda a^2\cos^2\theta)(1-\cos\theta)} X\label{eqn:general-Phi-k} \\
\tilde{\Phi}_k &= \frac{\left[ \frac{1}{\sqrt{2}} \lambda a(1 +\cos\theta) \left(U+V+ \sqrt{2}\Delta a\cos\theta\right)- 2  (1 +\lambda a^2\cos\theta)\right]}{2\mathcal{N}_k\Delta^2 a (1-\lambda a^2\cos^2\theta)(1+\cos\theta)}Y.\label{eqn:general-tilde-Phi-k}
\end{align}
It can be shown by a direct computation that $\Phi_\ell$ satisfies the non-linear PDEs (\ref{eqn:Klein-space-non-linear-PDEs}). 

These four functions are not independent, but are all related to each other by a sequence of discrete transformations. Although not obvious from the form written above, the functions $\tilde{\Phi}_\ell$ and $\tilde{\Phi}_k$ are identical. By inspection, we also have
\begin{align}
\begin{split}
\tilde{\Phi}_\ell(a,U,V,X,Y)&=\Phi_\ell(-a,U,V,Y,X), \\
\tilde{\Phi}_k (a,U,V,X,Y)&=-\Phi_k(a,-U,-V,Y,X),
\end{split}
\end{align}
 or in spheroidal coordinates
\begin{align}
\begin{split}
\tilde{\Phi}_\ell(a,t,r,\cos\theta,\varphi)&=\Phi_\ell(-a,t,r,\cos\theta,-\varphi), \\
\tilde{\Phi}_k (a,t,r,\cos\theta,\varphi)&=\Phi_k(a,-t,-r,\cos\theta,-\varphi).
\end{split}
\end{align}
The functions $\Phi_\ell$ and $\Phi_k$ which we will use for the Newman-Penrose map are then related by
\begin{align}
\Phi_\ell(a, t,r,\cos\theta,\varphi) = \Phi_k(-a,-t,-r,\cos\theta,\varphi).
\label{eqn:kelltrans}
\end{align}
It then follows that 
$\Phi_k$ satisfies the same non-linear PDEs (\ref{eqn:Klein-space-non-linear-PDEs}) as $\Phi_\ell$, while 
$\tilde{\Phi}_\ell=\tilde{\Phi}_k$ satisfies the non-linear PDEs obtained from (\ref{eqn:Klein-space-non-linear-PDEs}) by interchanging $X$ with $Y$. Each are therefore harmonic with respect to the flat metric, and for these double Kerr-Schild spacetimes the Newman-Penrose map can be defined. This means that the gauge field constructed from these scalar functions via the Newman-Penrose map will automatically  be a self-dual solution of the vacuum Maxwell equations in both flat and conformally flat backgrounds. The relation \eqref{eqn:kelltrans} also implies that the two parts of the gauge field are related via
\begin{align}
A_\ell (Q_e, a, t,r,\cos\theta, \varphi) &= A_k (Q_m, -a, -t, -r, \cos\theta, \varphi)
\label{eq:AellAk}
\end{align}
with $A_\ell =\hat{k}_e \Phi_\ell$ and $A_k = \hat{k}_m \Phi_k$.

\subsection{Self-dual gauge field}

To obtain the self-dual gauge field from this metric via the Newman-Penrose map, we act the operator $\hat{k}$ (with appropriate charges) defined in \eqref{eq:khatem} on the functions $\Phi_k$ and $\Phi_\ell$ from the previous section, i.e. 
\begin{align}
A_{\textnormal{NP}} &= \hat{k}_e \hat{\Phi}_\ell + \hat{k}_m \hat{\Phi}_k.
\label{eqn:NP-gauge-field}
\end{align}
We will set $\epsilon_0 = \mu_0 = 1$ from now on for ease of notation. In practice, we compute $A_\ell$ and use the relationship \eqref{eq:AellAk} to find $A_k$.

In order to compare with known results, we then analytically continue the gauge field to Lorentzian signature using $a\rightarrow - ia, \ \varphi \rightarrow i \varphi,\ Q_m \rightarrow  i Q_m$. Our total gauge field in Lorentzian signature is
  \begin{align}
  \begin{split}
A_{\textnormal{NP}} =&\  \frac{(Q_e+i Q_m)}{4\pi(r+ia\cos\theta) }\left[\left(1+\lambda iar \cos\theta\right)dt+dr+(r\cos\theta+ia) i d\varphi\right]  \\
&+A_{\textnormal{NP, gauge}}
\end{split}
 \end{align}
with the pure gauge term
   \begin{align}
   \begin{split}
A_{\textnormal{NP, gauge}} =&\  \frac{(Q_e+iQ_m)}{2\pi}\frac{\sqrt{2}}{4r \Omega}\frac{(1-\lambda r^2)}{(1+\lambda a^2)}  \bigg(2(1+\lambda a^2) +i \lambda\Delta a (r+i a) (1-\cos\theta)\bigg)\\
 &\ \ \ \ \times \left[dt-\frac{1}{(1-\lambda r^2)} dr-\frac{i a}{(1+\lambda a^2\cos^2\theta)}d(\cos\theta)\right]\\
 &-\frac{(Q_e+iQ_m)}{4\pi r}  \left[dt -i a \frac{1}{(1+\lambda a^2\cos^2\theta)}d(\cos\theta)\right]\\
&-\frac{(Q_e+iQ_m)}{2\pi} \frac{1}{2} i d\varphi -\frac{(Q_e+iQ_m)}{2\pi}\frac{d\Delta}{(2-\Delta)}\\
&-\frac{(Q_e+iQ_m)}{2\pi}\left[\frac{\sinh(\sqrt{\lambda} t)}{\cosh(\sqrt{\lambda} t)}\sqrt{\lambda} dt \right]+\frac{(Q_e+iQ_m)}{2\pi}\frac{1}{2r} \left[\frac{(1+\lambda r^2)}{(1-\lambda r^2)} dr\right]\\
&+\frac{(Q_e+iQ_m)}{2\pi}\left[\frac{(1-\lambda a^2\cos\theta)}{2 (1+\cos\theta)(1+\lambda a^2\cos^2\theta)}d(\cos\theta) \right]
\end{split}
 \end{align}
and 
\begin{align}
\Omega &= \frac{1}{\sqrt{2}}\bigg[2-\sqrt{\lambda} r (2-\Delta)\frac{\sinh(\sqrt{\lambda} t)}{\cosh(\sqrt{\lambda} t)}  - \lambda\Delta r^2  - \lambda\Delta a^2 (1-\cos\theta)\frac{(1-\lambda r^2)}{(1+\lambda a^2)}\bigg].
\end{align}
Finally, looking separately at the real and imaginary parts, we get

\begin{align}
\begin{split}
\Re(A_{\textnormal{NP}}) =&\  \frac{Q_e r }{4\pi(r^2+a^2\cos^2\theta) }\left[\left(1+\lambda a^2  \cos^2\theta\right)dt+dr-a\sin^2\theta  d\varphi\right] \\
&+\frac{ Q_m a\cos\theta }{4\pi(r^2+a^2\cos^2\theta) }\left[(1-\lambda r^2)dt+ dr-\frac{(r^2+a^2)}{a}  d\varphi\right] 
\label{totalreal}
\end{split}
 \end{align}
and
 \begin{align}
 \begin{split}
\Im(A_{\textnormal{NP}}) =&\  \frac{ Q_m r}{4\pi(r^2+a^2\cos^2\theta) }\left[(1+\lambda a^2\cos^2\theta) dt+dr-a\sin^2\theta d\varphi\right]\\
&-\frac{Q_e a\cos\theta}{4\pi(r^2+a^2\cos^2\theta) }\left[(1-\lambda r^2)dt+dr-\frac{(r^2+a^2)}{a}  d\varphi\right].
\end{split}
 \end{align}
For comparison, the Kerr-Schild double copy of a metric in double Kerr-Schild form \eqref{DKS} is defined as \cite{Luna:2015paa}
\begin{align}
A_{\textnormal{KS}} =  \psi \ell+\phi k.
\end{align}
For the Kerr-Taub-NUT-(A)dS metric in $(2,2)$ signature \eqref{KerrTaubNUTAdS}, we can perform a slightly modified version of the coordinate transformation \eqref{twistedspheroidalpq} ($t\rightarrow -t$) which does not change the background metric
\begin{align}
\begin{split}
dq &= dr\\
dp &= d(a \cos\theta)\\
d\tilde{\tau} &= \frac{1}{(1-\lambda a^2)}\left(-dt-\frac{1}{(1-\lambda r^2)}dr +ad\varphi\right) - \frac{a \cos^2\theta}{(1-\lambda a^2 \cos^2\theta)\sin\theta} d\theta \\
d\tilde{\sigma} &= \frac{1}{(1-\lambda a^2)}\left(\lambda dt+ \frac{\lambda}{(1-\lambda r^2)}dr - \frac{1}{a}d\varphi\right)+ \frac{1}{(1-\lambda a^2 \cos^2\theta)a\sin\theta} d\theta.
\end{split}
\label{twistedspheroidalpq2}
\end{align}
In these coordinates the two scalar functions $\phi$ and $\psi$ are
\begin{align}
\psi &= \frac{2Mq}{q^2-p^2} =\frac{2Mr}{(r^2-a^2\cos^2\theta)}\\
\phi &= \frac{2Np}{q^2 - p^2} = \frac{2Na\cos\theta}{(r^2 - a^2\cos^2\theta)},
\end{align}
and the vectors are 
\begin{align}
\ell &=  \frac{(1-\lambda a^2\cos^2\theta)}{(1-\lambda a^2)}dt +\frac{(1-\lambda a^2\cos^2\theta)}{(1-\lambda r^2)(1-\lambda a^2)} dr -\frac{a\sin^2\theta}{(1-\lambda a^2)}d\varphi\\
k &= \frac{(1- \lambda r^2)}{(1- \lambda a^2)} dt + 
 \frac{1}{(1- \lambda a^2)} dr - \frac{(r^2-a^2 \cos^2 \theta)}{a \sin \theta 
 (1- \lambda a^2 \cos^2 \theta)} d \theta - \frac{(r^2-a^2)}{a(1-\lambda a^2)}d \varphi.
\end{align}
 After analytic continuation to Lorentzian signature ($a\rightarrow - ia, \ \varphi \rightarrow i \varphi,\ N \rightarrow  i N$), the Kerr-Schild gauge field for this metric is then  (mapping $M \rightarrow Q_e/8\pi$ and $N \rightarrow Q_m/8\pi $) 
\begin{align}
\begin{split}
A_{\textnormal{KS}} 
=&\ \frac{Q_e r}{4\pi (r^2+a^2\cos^2\theta)(1+\lambda a^2)}\bigg[(1+\lambda a^2\cos^2\theta)dt + dr -a\sin^2\theta d\varphi\bigg]\\
&+ \frac{ Q_m a \cos\theta}{4\pi (r^2+a^2\cos^2\theta)(1+\lambda a^2)}\bigg[(1- \lambda r^2) dt + 
 dr  - \frac{(r^2+a^2)}{a}d \varphi\bigg]\\
 &+A_{\textnormal{KS, gauge}} 
 \end{split}
\end{align}
with
\begin{align}
A_{\textnormal{KS, gauge}} 
&= \frac{Q_e}{4\pi (1+\lambda a^2)}\bigg[\frac{\lambda r}{(1-\lambda r^2)} dr \bigg]- \frac{ i Q_m }{4\pi }\bigg[  \frac{\cos\theta}{  \sin \theta 
 (1+ \lambda a^2 \cos^2 \theta)} d \theta \bigg].
\end{align}
This agrees exactly with the real part of the Newman-Penrose map \eqref{totalreal} up to an overall constant factor of $\frac{1}{(1+\lambda a^2)}$ and pure gauge terms.

In the $\lambda \rightarrow 0$ limit the metric is the Kerr-Taub-NUT solution, and the gauge field becomes
\begin{align}
\begin{split}
\Re(A_{\textnormal{NP}})\bigg|_{\lambda \rightarrow 0} =&\  \frac{Q_e r }{4\pi(r^2+a^2\cos^2\theta) }\left[dt+dr-a\sin^2\theta  d\varphi\right] \\
&+\frac{ Q_m a\cos\theta }{4\pi(r^2+a^2\cos^2\theta) }\left[dt+ dr-\frac{(r^2+a^2)}{a}  d\varphi\right] 
\end{split}
 \end{align}

 \begin{align}
 \begin{split}
\Im(A_{\textnormal{NP}})\bigg|_{\lambda \rightarrow 0} =&\  \frac{  Q_m r}{4\pi(r^2+a^2\cos^2\theta) }\left[ dt+dr-a\sin^2\theta d\varphi\right]\\
&-\frac{Q_e a\cos\theta}{4\pi(r^2+a^2\cos^2\theta) }\left[dt+dr-\frac{(r^2+a^2)}{a}  d\varphi\right].
\end{split}
 \end{align}
 The electric part of this solution in this limit is exactly equivalent to the one found in \cite{Elor:2020nqe} (known as $\sqrt{\textnormal{Kerr}}$) for the Kerr metric, and the addition of the NUT charge simply adds the magnetic dual part. The electric part of this gauge field was shown in \cite{Monteiro:2014cda, Arkani-Hamed:2019ymq} to be sourced by an axisymmetric electric charge
distribution rotating at a uniform rate about the $z$-axis (where $z = r\cos\theta$). We expect the magnetic part of this field to be similarly sourced by an axisymmetric magnetic charge
distribution rotating at a uniform rate about the $z$-axis.

In the $a\rightarrow 0$ limit, the gravitational solution is the Taub-NUT-(A)dS metric. The gauge solution can be written (with the addition/subtraction of pure gauge terms) as 
 \begin{align}
\Re(A_{\textnormal{NP}})\bigg|_{a\rightarrow 0} =&\  \frac{Q_e }{4\pi r }dt  +\frac{ Q_m }{4\pi }(1-\cos\theta ) d\varphi \\
\Im(A_{\textnormal{NP}})\bigg|_{a\rightarrow 0} =&\  \frac{ Q_m }{4\pi r }dt -\frac{Q_e }{4\pi }(1-\cos\theta)  d\varphi.
 \end{align}
 The real part of this solution is a dyon (a Coulomb charge plus a magnetic monopole). This is equal to the gauge solution found via the Kerr-Schild double copy in \cite{Luna:2015paa} in flat space. The electric part of this solution also agrees with the Schwarzschild-(A)dS solution (up to pure gauge terms) found via the Kerr-Schild double copy in \cite{Bahjat-Abbas:2017htu, Carrillo-Gonzalez:2017iyj}. 
 
 In this limit, the $\lambda$ dependence of the solution completely drops out. Although a priori in the $a\rightarrow 0$ limit the $\Phi$'s appear to depend on $\lambda$ through $\Delta$, this dependence is completely eliminated upon expressing $\Phi$'s in terms of null coordinates ($U,V, X,Y$) only. For the $\Phi$'s depend on $r$ and $\Delta$ only in the combination $\Delta r$, and it is only this combination that appears in the quadratic equation relating $r$ to the null coordinates.

In other words, 
\begin{align}
    \Phi(a=0, \lambda, t,r, \cos \theta, \varphi)&=\Phi(a = 0, \lambda = 0,U,V,X,Y). 
\end{align}
is independent of $\lambda$. Again, this result isn't too surprising given that $\Phi$ describes a shear-free null geodesic congruence, physical properties that are conformally invariant.

 \subsection{Electric-magnetic duality}

The field strength $F_{\textnormal{NP}}=dA_{\textnormal{NP}}$ is of the form 
\begin{align}
F_{\textnormal{NP}}= (Q_e +iQ_m)F 
\end{align}

where both $F_{\textnormal{NP}}$ and $F$ are self-dual, with $F_{\textnormal{NP}}=i\star_0 F_{\textnormal{NP}}$ and $F=i \star_0F$. 
This solution exhibits a discrete electric-magnetic duality that acts on the charges, separate from its Hodge star self-dual property.
From the self-dual property of $F$ it follows that  
\begin{align}
    F= f + i \star_0 f
\end{align}
for some real two-form $f$, and also that
\begin{align}
    F_{\textnormal{NP}} = Q_e f -Q_m \star_0 f  +i(Q_m f+Q_e \star_0 f).
\end{align}

This result implies a relation between the complex electric $E$ and magnetic $B$ fields (vector notation suppressed) of $F_{\textnormal{NP}}$ and the real-valued electric $E_f$ and magnetic $B_f$ components of $f$, namely 
\begin{align}
\begin{split}
\Re{(E)} &=Q_e E_f - Q_m B_f  \\
\Re{(B)} &=Q_e B_f + Q_m E_f  \\
\Im{(E)} &=Q_m E_f + Q_e B_f  \\
\Im{(B)} &=Q_m B_f - Q_e E_f 
\end{split}
\end{align}
where $\star E=B$ and $\star B=-E$ has been used. Comparing these equations gives 
$\Im{(E)}=\Re{(B)}$ and $\Im{(B)}=-\Re{(E)}$, which taken together is none other than the statement that $F_{\textnormal{NP}}$ is a self-dual solution, i.e., $E = i B$. Note however that $\Re{(E)}$ and $-\Re{(B)}$, which are not related by the Hodge star operation,  are obtained from $\Re{(B)}$ and $\Re{(E)}$, respectively,  by $(Q_e, Q_m) \rightarrow (-Q_m,Q_e)$, which is a counter-clockwise $\mathit{U}(1)$ rotation on the electric and magnetic charge by an angle $\theta = \pi/2$.

Given such a self-dual $F_{\textnormal{NP}}$, one can take $\Re{(F_{\textnormal{NP}})}$ alone (or $\Im{(F_{\textnormal{NP}})}$), forget the other, and obtain a non-self dual solution to the source-free Maxwell's equation. The resulting field strength will also exhibit electric-magnetic duality.

For gravitational solutions on a flat background with vanishing rotational parameter $a$, the object on the gauge field side describes a dyon which has manifest electric-magnetic duality.  Here we find this electric-magnetic duality of the solution carries over to maximally symmetric spacetimes, including those having non-vanishing rotational parameter. 
The gravitational analog of electric-magnetic duality has been explored in a general setting using the Kerr-Taub-NUT spacetime as a case study in \cite{Kol:2022bsd}.

\section{Discussion}

In this paper we applied the intuition of the Newman-Penrose map to more general spacetimes, including those in double Kerr-Schild form and on maximally symmetric backgrounds.
Although we do not have a rigorous formulation of the Newman-Penrose map for general double Kerr-Schild metrics, for a particular example metric we are able to reproduce the gauge field from the Kerr-Schild and Weyl double copies, as shown in table \ref{table:main-table}.

\begin{table}[h!]
\begin{center}
\begin{tabular}{|c| c| c| c |c |} 
 \hline
 Gravity solution & Gauge solution & KS/Weyl DC  & NP Map \\ 
 \hline\hline
 Schwarzschild & Coulomb &  \cite{Monteiro:2014cda}  &\cite{Elor:2020nqe} \\ 
 \hline
 Kerr & rotating disc of charge & \cite{Monteiro:2014cda} & \cite{Elor:2020nqe} \\ \hline
 Photon Rocket & Lienard-Weichert potential & \cite{Luna:2016due} & \cite{Elor:2020nqe} \\ \hline
Taub-NUT & dyon & \cite{Luna:2015paa} & this work \\ \hline
Kerr-Taub-NUT & rotating dyon & \cite{Luna:2015paa,Bahjat-Abbas:2020cyb} & this work\\ \hline
Schwarzschild-(A)dS & Coulomb in (A)dS & \cite{Bahjat-Abbas:2017htu, Carrillo-Gonzalez:2017iyj}  & this work\\
 \hline
 Kerr-(A)dS & rotating disc in (A)dS & \cite{Carrillo-Gonzalez:2017iyj} & this work\\ \hline 
 Kerr-Taub-NUT-(A)dS & aligned fields in (A)dS & \cite{Chawla:2022ogv} & this work \\ \hline 
\end{tabular}
\caption{Table enumerating example metrics and their gauge fields obtained via either the Kerr-Schild/Weyl double copy or the Newman-Penrose map.}
\label{table:main-table}
\end{center}
\end{table}

In particular, we studied the Kerr-Taub-NUT-(A)dS metric which can be put into double Kerr-Schild form after analytic continuation to Kleinian signature. In analogy with the Newman-Penrose map for single Kerr-Schild spacetimes with a flat background, we obtained scalar functions $\Phi_\ell$ and $\Phi_k$ for each of the null vectors by identifying null coordinates that make the conformal flatness and Lorentz invariance of the maximally symmetric background metric manifest. These scalar functions are shown to satisfy the non-linear PDEs sufficient for them to be harmonic in flat spacetime (but not conformally flat spacetime).  

Acting with our spin-raising operator gives a gauge field, which, because of the harmonic conditions on the $\Phi_\ell$ and $\Phi_k$, satisfies the vacuum Maxwell equations in \emph{both} flat and conformally flat spacetime. We then compare our results with those of the Kerr-Schild and Weyl double copies and show they match up to pure gauge terms. We also find that the field strength obtained from the Newman-Penrose map gauge field displays a discrete electric-magnetic duality.

The fact that $\Phi_\ell$ satisfies the non-linear PDEs implies it be can be obtained 
from the zero set of a holomorphic function $F$ (the ``Kerr function'') of twistor variables, one of the consequences of Kerr's Theorem \cite{Penrose:1967wn,Penrose:1968me,osti_7216328,1976CMaPh..47...75C}. What this function is in general for the Kerr-Taub-NUT-(A)dS family of solutions is presently unknown. In the $a=0, \lambda \neq 0$ limit it is seen to be identical to the function $F_{Sch}$ of the Schwarzschild spacetime, and for $\lambda=0$ it is a quadratic function of $\Phi_\ell$ and invariants of the non-linear PDEs (and obtainable from $F_{Sch}$ upon applying the Newman-Janis trick). Knowing that the Kerr function is holomorphic in twistor variables is insufficient information to fully fix it. For more general metrics of the single Kerr-Schild metric form, the Kerr function is obtained by integrating the Einstein equations \cite{1969JMP....10.1842D} (see also,\cite{mcintosh1988single}). Finding the Kerr function in general for double Kerr-Schild metrics, or even for the subset of the Kerr-Taub-NUT-(A)dS family of solutions, presumably involves similar steps.

Our results suggest that the Newman-Penrose map might be generalizable to all double Kerr-Schild spacetimes admitting an integrable distribution of totally null vectors. For flat backgrounds, these can all be solved exactly \cite{1976NCimB..35...35P}. The tetrad can be generalized to accommodate double Kerr-Schild spacetimes, but this only leads to a subset of the non-linear PDEs being satisfied, so there remains a conceptual obstruction to defining the Newman-Penrose map for general double Kerr-Schild spacetimes.

These results provide further evidence that the Newman-Penrose map is a manifestation of the double copy, with the potential to generically study non-vacuum and non-stationary solutions. In future work we hope to understand the relationship between the Newman-Penrose map and the Kerr-Schild double copy, which a priori appear very different, yet agree in all known examples. This promises to provide insight into the structure of the double copy more generally, and the gravitational and gauge theories to which it applies. The apparently close relationships between the Kerr-Schild and Weyl double copies and the Newman-Penrose map also serve to highlight a shortcoming that all three constructions share---namely that they map exact gravitational solutions to \emph{abelian} gauge fields, while the presumably more fundamental double copy for amplitudes necessarily relates graviton amplitudes with \emph{gluon} amplitudes. Generalizing the Newman-Penrose map to the non-abelian case could potentially illuminate the structures underlying both classical and quantum mechanical double copies.

\section*{Acknowledgements}

 The work of KF is supported by the Swiss National Science Foundation 
 under grant no. 200021-205016. KF also acknowledges support from
Simons Foundation Award Number 658908. The work of MG is supported by the LDRD program at Los Alamos National Laboratory and by the U.S. Department of Energy, Office of High Energy Physics, under Contract No. DE-AC52-06NA25396. The Los Alamos preprint number for this 
manuscript is LA-UR-23-26863.


\newpage
\bibliographystyle{JHEP}
\bibliography{bib}

\end{document}